\documentstyle[12pt,a4]{article}
\newcommand{\vf}{\varphi}
\newcommand{\be}{\begin{equation}}
\newcommand{\ee}{\end{equation}}
\newcommand{\ba}{\begin{eqnarray}}
\newcommand{\ea}{\end{eqnarray}}

\newcommand{\bi}{\bibitem}
\begin{document}
\bigskip\bigskip\begin{center}
{\bf \Large{GENERALIZED KILLING EQUATIONS AND \\
~\\                                                                  
TAUB-NUT SPINNING SPACE}}
\end{center}
\vskip 1.0truecm
\centerline{{\bf\large
{Diana Vaman\footnote{E-mail address:~~~~DVAMAN@THEOR1.IFA.RO}}~~    
{\it and}~~ {\bf\large Mihai Visinescu\footnote
{E-mail address:~~~ MVISIN@THEOR1.IFA.RO}}}}
\vskip5mm
\centerline{Department of Theoretical Physics}
\centerline{Institute of Atomic Physics, P.O.Box MG-6, Magurele,}
\centerline{Bucharest, Romania}                                      
\vskip 2cm
\bigskip \nopagebreak \begin{abstract}
\noindent

The generalized Killing equations for the configuration space of
spinning particles (spinning space) are analysed.
Simple solutions of the homogeneous part of these equations are
expressed in terms of Killing-Yano tensors. The general
results are applied to the case of the four-dimensional
euclidean Taub-NUT manifold.
\end{abstract}

~~~~~02.40.+m Geometry, differential geometry, and topology
\vskip 2cm
\section{Spinning particles and constants of motion}
The pseudo-classical limit of the Dirac theory of a spin 1/2
fermion in curved space-time is described by the supersymmetric
extension of the usual relativistic point-particle [1]. The
configuration space of spinning particles (spinning space) is an
extension of an ordinary Riemannian manifold, parametrized by
local coordinates $\left\lbrace x^\mu \right\rbrace$, to a graded 
manifold parametrized
by local coordinates $\left\lbrace x^\mu,\psi^\mu\right\rbrace$, 
with the first set of
variables being Grassmann-even (commuting) and the second set
Grassmann-odd (anti-commuting).
The equation of motion of a spinning particle on a geodesic is
derived from the action:
\be
S=\int d\tau\left( \frac12 g_{\mu\nu}(x)\dot{x}^\mu \dot{x}^\nu + 
\frac{i}{2} 
g_{\mu\nu}(x)\psi^\mu\frac{D\psi^\nu}{D\tau}\right) .
\ee

The corresponding world-line hamiltonian is given by:
\be 
H=\frac1 2 g^{\mu\nu}\Pi_\mu \Pi_\nu
\ee
where $\Pi_\mu = g_{\mu\nu}\dot{x}^\nu$ is the covariant momentum. 

For any constant of motion ${\cal J}(x,\Pi,\psi)$, the bracket
with $H$ vanishes,
$\left\lbrace H,{\cal J} \right\rbrace = 0$,
with the Poisson-Dirac brackets for functions of the covariant       
phase-space variables $(x,\Pi,\psi)$ defined by
\be
\left\lbrace F,G\right\rbrace={\cal D}_\mu F\frac{\partial G}
{\partial
\Pi_\mu} - \frac{\partial F}{\partial \Pi_\mu}{\cal D}_\mu G -
{\cal R}_{\mu\nu}\frac{\partial F}{\partial
\Pi_\mu}\frac{\partial G}{\partial \Pi_\nu} +
i(-1)^{a_F}\frac{\partial F}{\partial \psi^\mu}\frac{\partial
G}{\partial \psi_\mu}
\ee
where the notations used are
\be
{\cal D}_\mu F = \partial_\mu +
\Gamma^\lambda_{\mu\nu}\Pi_\lambda\frac{\partial F}{\partial
\Pi_\nu} - \Gamma^\lambda_{\mu\nu}\psi^\nu \frac{\partial
F}{\partial \psi^\lambda}~~~;~~~
{\cal R}_{\mu\nu} = \frac{i}{2}\psi^\rho\psi^\sigma 
R_{\rho\sigma\mu\nu}
\ee
and $a_F$ is the Grassmann parity of $F$ : $a_F=(0,1)$ 
for $F$=(even,odd).
\par
If we expand ${\cal J}(x,\Pi,\psi)$ in a power series in the
canonical momentum 
\be
{\cal J}=\sum_{n=0}^{\infty}\frac{1}{n!}{\cal J}^{(n)\mu_1
\dots\mu_n}(x,\psi) \Pi_{\mu_1}\dots\Pi_{\mu_n}
\ee
then the bracket $\{ H , {\cal J}\}$ vanishes for
arbitrary $\Pi_\mu$ if and only if the components of ${\cal J}$ 
satisfy the generalized Killing equations [1] :
\be
{\cal J}^{(n)}_{(\mu_1\dots\mu_n;\mu_{n+1})} + \frac{\partial
{\cal J}^{(n)}_{(\mu_1 \dots\mu_n}}{\partial \psi^\sigma}
\Gamma^\sigma_{\mu_{n+1})\lambda} \psi^\lambda = 
\frac{i}{2}\psi^\rho \psi^\sigma R_{\rho\sigma\nu(\mu_{n+1}}
{{\cal J}^{(n+1)\nu}}_{\mu_1 \dots \mu_n)}
\ee
where the parantheses denote full symmetrization over the indices    
enclosed.

The solutions of the generalized Killing equations (6) can be
divided into two clases [2,3]: {\it generic} ones, which exists
for any spinning particle model (1) and {\it non-generic} ones,
which depend on the specific background space considered. To the
first class belong proper-time translations and supersymmetry,
generated by the hamiltonian and supercharge:
\be
Q_0=\Pi_\mu \psi^\mu.
\ee

In addition there is also a "chiral" symmetry generated by the
chiral charge
\be
\Gamma_* = \frac{i^{[\frac{d}{2}]}}{d!}\sqrt{g}\epsilon_{\mu_1 
\dots \mu_d} \psi^{\mu_1} \dots \psi^{\mu_d}
\ee
and a dual supersymmetry whose generator is
\be
Q^* = i\{ \Gamma_* , Q_0 \} = 
\frac{i^{[\frac{d}{2}]}}{(d-1)!}\sqrt{g}\epsilon_{\mu_1 \dots 
\mu_d} \Pi^{\mu_1}\psi^{\mu_2} \dots \psi^{\mu_d}
\ee
where $d$ is the dimension of space-time.

The {\it non-generic} conserved quantities depend on the
explicit form of the metric $g_{\mu\nu}(x)$. It was a great
success of Gibbons et al.[3] to have been able to prove that the
Killing-Yano tensors can be understood as objects generating
{\it non-generic} supersymmetries. A tensor $f_{\mu_1 \dots\mu_r}$ 
is called Killing-Yano of valence $r$ if it is totally antisymmetric 
and satisfies the equation  
\be
f_{\mu_1 \dots\mu_{r-1}(\mu_{r};\lambda)} = 0.
\ee

In order to solve the system of coupled differential equations
(6) one starts with a $\tilde {\cal J}^{(n)}_{\mu_1 \dots
\mu_n}$ solution of the homogeneous equation:
\be
\tilde{\cal J}^{(n)}_{(\mu_1 \dots\mu_n;\mu_{n+1})}+
\frac{\partial \tilde{\cal J}^{(n)}_{(\mu_1 \dots\mu_n}}
{\partial \psi^\sigma}
\Gamma^\sigma_{\mu_{n+1})\lambda}\psi^\lambda = 0.
\ee

This solution is introduced in the r.h.s. of the generalized
Killing equation (6) for ${\cal J}^{(n-1)}_{\mu_1
\dots\mu_{n-1}}$ and the iteration is carried on to $n=0$.

In fact, for the bosonic sector, neglecting the Grassmann
variables $\{ \psi^\mu \}$, all the generalized Killing
equations (6) are homogeneous and decoupled. The first equation
shows that ${\cal J}_0$ is a trivial constant, the next one is
the equation for the Killing vectors and so on. In general, the
homogeneous equation for a given $n$ defines a Killing tensor of
valence $n$ and ${\cal J}^{(n)}_{\mu_1
\dots\mu_n}\Pi^{\mu_1}\dots\Pi^{\mu_n}$ is a first integral of
the geodesic equation [4].

For the spinning particles, even if one starts with a Killing
tensor of valence $n$, solution of eq.(11) in which all spin
degrees of freedom are neglected, the components 
${\cal J}^{(m)}_{\mu_1\dots\mu_m}~~~~(m<n)$ will receive a 
nontrivial spin contribution. 

In what follows we should like to stress that the very starting
homogeneous equation (11) can have solutions depending on the
Grassmann coordinates $\{ \psi^\mu \}$. That is the case of the 
manifolds admitting Killing-Yano tensors. For example, for the first 
equation (11), i.e. $n=0$,
\be
\tilde {\cal J}^{(0)} = \frac i4 f_{\mu\nu}\psi^\mu\psi^\nu
\ee
is a solution if $f_{\mu\nu}$ is a Killing-Yano tensor
covariantly constant. Moreover $\tilde{\cal J}^{(0)}$ is a
separately conserved quantity.

Going to the next equation (11) with $n=1$, a natural solution
is: 
\be
\tilde{\cal J}^{(1)}_\mu = R_\mu f_{\lambda\sigma}\psi^\lambda
\psi^\sigma
\ee
where $R_\mu$ is a Killing vector ($R_{(\mu;\nu)} = 0$) and
again $f_{\lambda\sigma}$ is a Killing-Yano tensor covariantly
constant. Introducing this solution in the r.h.s. of the eq. (6)
with $n=0$, after some calculations, we get for ${\cal J}^{(0)}$
\be
{\cal J}^{(0)} = \frac i2 R_{[\mu;\nu]}f_{\lambda\sigma}\psi^\mu 
\psi^\nu \psi^\lambda \psi^\sigma
\ee
where the square bracket denotes the antisymmetrization with
norme one. Finally, from eq.(5) with the aid of eqs.(13) and
(14) we get a new constant of motion which is peculiar to the
spinning case:
\be
{\cal J} = f_{\mu\nu}\psi^\mu \psi^\nu \left( R_\lambda
\Pi^\lambda + \frac i2 R_{[\lambda;\sigma]}\psi^\lambda \psi^\sigma
\right).
\ee

Another $\psi$-dependent solution of the $n=1$ eq.(11) can be
generated from a Killing-Yano tensor of valence $r$:
\be
\tilde{\cal J}_{\mu_1}^{(1)} = f_{\mu_1 \mu_2\dots\mu_r}\psi^{\mu_2}
\dots \psi^{\mu_r}.
\ee

Following the above prescription we get for ${\cal J}^{(0)}$:
\be
{\cal J}^{(0)} = \frac{i}{r+1}(-1)^{r+1} f_{[\mu_1\dots
\mu_r;\mu_{r+1}]} \psi^{\mu_1}\dots\psi^{\mu_{r+1}}
\ee
and the constant of motion corresponding to these solutions
of the Killing equations is:
\be
Q_f = f_{\mu_1 \dots\mu_r}\Pi^{\mu_1}\psi^{\mu_2}\dots \psi^{\mu_r} 
+ \frac{i}{r+1}(-1)^{r+1}f_{[\mu_1 \dots
\mu_r;\mu_{r+1}]}\psi^{\mu_1}\dots \psi^{\mu_{r+1}}.
\ee

Therefore the existence of a Killing-Yano tensor of valence $r$ 
is equivalent to the existence of a supersymmetry for the
spinning space with supercharge $Q_f$ which anticommutes with
$Q_0$. A similar result was obtained in ref.[5] in which it is
discussed the role of the generalized Killing-Yano tensors, with
the framework extended to include electromagnetic interactions.

Finally, we should like to mention the special case of a covariantly 
constant tensor ${\cal J}^{(n)}_{\mu_1 \dots\mu_n}$, symmetric in 
the first $r$ indices and antisymmetric in the remaining ones. Using
such kind of tensor, the Killing 
equations are decoupled even in the spinning case, 
the quantity ${\cal J}^{(n)}_{\mu_1\dots\mu_n}
\Pi^{\mu_{1}}\dots\Pi^{\mu_{r}} 
\psi^{\mu_{r+1}}\dots \psi^{\mu_{n}}$ being
conserved along the geodesics.

In the main, with some ability, it is possible to investigate
higher orders of eq.(11), but it seems that one cannot go too
far with simple, transparent expressions. Instead of that, we
shall apply the above constructions to a concret case, namely
the four-dimensional euclidean Taub-NUT manifold.

\section{Taub-NUT spinning space}
Much attention has been paid to the Euclidean Taub-NUT metric, 
since in the long distance limit the relative motion of two monopoles 
is described approximately by its geodesics [6,7]. As it is well
known, the geodesic motion of the Taub-NUT metric admits the 
Kepler-type symmetry [8-11]. On the other hand, the
Kaluza-Klein monopole of Gross and Perry [12] and Sorkin [13]
was obtain by embedding the Taub-NUT gravitational instanton
into five-dimensional Kaluza-Klein theory.

In a special choice of coordinates the euclidean Taub-NUT metric
takes the form 
\be
ds^2 = V(r)\left(dr^2+r^2 d\theta^2+r^2\sin^2\theta
 d\varphi^2\right) + 16m^2 V^{-1}(r)(d\chi+\cos\theta
 d\varphi)^2
\ee
with $V(r)=1 + \frac{4m}{r}$.
There are four Killing vectors [8-11]
\be
D_{A}=R_{A}^\mu\,\partial_\mu,~~~~A=0,\cdots ,3
\ee
corresponding to the invariance of the metric (19) under spatial 
rotations $(A=1,2,3)$ and $\chi$ translations $(A=0)$.
In the purely bosonic case these invariances would correspond to 
conservation of angular momentum and "relative electric charge" 
[8-10]:
\be
\vec{j}=\vec{r}\times\vec{p}\,+\,q\,{\vec{r}\over r} .
\ee
\be
q=16m^2\,V(r)\,(\dot\chi+\cos\theta\,\dot\vf) 
\ee
where $\vec{p}={1\over V(r)}\dot{\vec{r}}$ is the 
"mechanical momentum" which is only part of the momentum
canonically conjugate to $\vec{r}$.  

Finally, there is a conserved vector analogous to the
Runge-Lenz vector of the Kepler-type problem
\be
\vec{K} = \frac 12 \vec{K}_{\mu\nu}\dot{x}^\mu\dot{x}^\nu = 
\frac 12 \left[ \vec{p}\times\vec{j} + \left(\frac{q^2}{4m}-
4mE\right)\frac{\vec r}{r}\right]
\ee
where the conserved energy $E$, from eq.(2), is 
\be
E={1\over 2}\, g^{\mu\nu}\,\Pi_\mu\,\Pi_\nu\,
={1\over 2}V^{-1}(r)\left[\dot{\vec{r}}^{\,2} +\left( {q\over 
4m}\right)^2\right].
\ee

In the Taub-NUT geometry there are known to exist four
Killing-Yano tensors [9].
The first three Killing-Yano tensors $f_{i\mu\nu}$ are
covariantly constants (with vanishing field strength)
\be
f_i = 8m(d\chi + \cos\theta d\varphi)\wedge dx_i -
\epsilon_{ijk}(1+\frac{4m}{r}) dx_j \wedge dx_k.
\ee

The fourth Killing-Yano tensor is
\be 
f_Y = 8m(d\chi + \cos\theta  d\varphi)\wedge dr +
4r(r+2m)(1+\frac{r}{4m})\sin\theta  d\theta \wedge d\varphi
\ee
and having only one non-vanishing component of the field strength
\be
{f_{Y}}_{r\theta;\varphi} = 2(1+\frac{r}{4m})r\sin\theta.
\ee

The corresponding supercharges (18) constructed from the 
Killing-Yano tensors (25) and (26) are $Q_i$ and $Q_Y$. The
supercharges $Q_i$ together with $Q_0$ from eq.(7) realize the
N=4 supersymmetry algebra [14]:
\be
\left\{ Q_A , Q_B \right\} = -2i\delta_{AB}H~~~,~~~A,B=0,\dots,3
\ee
making manifest the link between the existence of the
Killing-Yano tensors and the hyper-K\"ahler geometry of the
Taub-NUT manifold.

Starting with these results from the bosonic sector of the
Taub-NUT space one can proceed with the spin contributions. The
first generalized Killing equation (6) shows that with each
Killing vector $R_{A}^{\mu}$ (20) there is an associated Killing
scalar $B_A$ [15]. A simple expression for the Killing scalar
was given in Ref.[14]:
\be
B_{A} = \frac i2 R_{A[\mu;\nu]}\psi^\mu \psi^\nu.
\ee

Therefore the total angular momentum and "relative electric
charge" become in the spinning case
\ba 
{\vec J}&=& {\vec B} + {\vec j}\\
J_0 &=& B_0 + q
\ea
where ${\vec J} =(J_1, J_2, J_3)$ and ${\vec B} =
(B_1, B_2, B_3)$.

The above constants of motion are superinvariant:
\be
\left\{ J_A , Q_0 \right\} = 0~~~;~~~A=0,\dots,3.
\ee

The Lie algebra defined by the Killing vectors is realized by
the constants of motion (30), (31) through the Poisson-Dirac
brackets (3).

Similarly, introducing the Killing tensors ${\vec K}_{\mu\nu}$ (23)
into the r.h.s. of the second generalized Killing equation (6)
we get that the corresponding Killing vectors ${\cal{\vec R}}_\mu$
have a spin dependent part ${\vec S}_\mu$ [16]
\be
{\cal \vec{R}}_\mu = \vec{R}_\mu + \vec{S}_\mu
\ee 
where $\vec{R}_\mu$ are the standard Killing vectors. The
$\psi$-dependent part of the Killing vectors $\vec{S}_\mu$
contribute to the Runge-Lenz vector for the spinning space 
\be
\vec{{\cal K}} =\frac 12 \vec{K}_{\mu\nu}\cdot\dot{x}^\mu\dot{x}^\nu
+  \vec{S}_\mu \cdot\dot{x}^\mu.
\ee

In terms of the supercharges $Q_i$ and $Q_Y$, the components of
the Runge-Lenz vector $\vec{\cal K}$ are given by [14]
\be
\vec{\cal K}_{i} = i \left\{ Q_Y , Q_{i} \right\}
~~~,~~~{i}=1,2,3.
\ee

The non-vanishing Poisson brackets are (after some algebra):
\ba
\left\{ J_{i} , J_{j} \right\} &=& \epsilon_{ijk} J_k\\
\left\{ J_{i} , {\cal K}_{j}\right\} &=& \epsilon_{ijk}
{\cal K}_{k}\\
\left\{ {\cal K}_{i} , {\cal K}_{j} \right\} &=&
\left( \frac{J_{0}^2}{16 m^2} -
2E \right) \epsilon_{ijk}J_{k}
\ea
similarly to the results from the bosonic sector [9].

Taking into account the existence of the Killing-Yano
covariantly constants tensors ${f_{i}}_{\mu\nu}$ (25), three
constants of motion can be obtained using the prescription (12):
\be
S_{i} = \frac i4 f_{i\mu\nu}\psi^\mu \psi^\nu
~~~,~~~ i=1,2,3
\ee
which realize an $SO(3)$ Lie-algebra similar to that of the
angular momentum (36):
\be
\left\{S_{i} , S_{j}\right\} = \epsilon_{ijk}
S_{k}.
\ee

These components of the spin are separately conserved and can be
combined with the angular momentum $\vec{J}$ to define a new
improved form of the
angular momentum $I_{i} = J_{i} - S_{i}$ with the property that it 
preserve the algebra 
\be
\left\{ I_{i},I_{j} \right\} = 
\epsilon_{ijk}I_{k}
\ee
and that it commutes with the $SO(3)$ algebra generated by the
spin $S_{i}$ 
\be
\left\lbrace I_{i},S_{j} \right\rbrace = 0.
\ee

Let us note also the following Dirac brackets of $S_{i}$ with 
supercharges
\be
\left\{ S_{i},Q_0 \right\} = -\frac{Q_{i}}{2}~~~~;~~~
\left\{ S_{i},Q_{j} \right\} = \frac1 2 (\delta_{ij} Q_0
+ \epsilon_{ijk}Q_{k}).
\ee

We can combine these two $SO(3)$ algebras (40), (41) to obtain
the generators of a conserved $SO(4)$ symmetry among the
constants of motion, a standard basis for which is spanned by
$M_{i}^\pm = I_{i} \pm S_{i}$ [14].
We should like to remark that there is no spin component like in
equation (39) to be used for a improved "relative electric
charge" $J_0$. The reason is that the fourth Killing-Yano
tensor $f_Y$ (26) is not covariantly constant. Of course, we
can add to $J_0$ a combination of the three separately
conserved quantities $S_{i}$ but this is not a natural "improved
relative electric charge". Moreover, it is impossible to modify a
particular solution $J_0$ (31) of the generalized Killing eq.(6)
for $n=0$ and Killing vectors in the r.h.s. by adding solutions of 
the homogeneous part of this equation in order to recover the
conserved quantity (22) from the standard Taub-NUT case as it
was suggested in Ref.[17].

In fact, from equation (31), in the spinning case, $q$ is not
separately conserved and we have [15]:
\be
\dot{q}(J_0 -q) = \dot{q}B_0 = -B_0\dot{B}_0 =
\frac{256 m^4}{(4m+r)^5}\dot{r}\Gamma_{*}.
\ee

Therefore, $q$ is not separately conserved in the spinning
space excepting the case $\Gamma_* = 0$. $\Gamma_*$ can be zero
if there is a relation between Grassmann variables ${\psi^\mu}$.
Such a relation can be realize impossing, for example, $Q_0 =
0$. This constraint can be correlated with the absence of an
intrinsic electric dipole moment of physical fermions (leptons
and quarks) [1]. The conservation of $Q_0$ guarantees that this
condition can be satisfied at all times, irrespective of the
presence of external fields.

However, in general, the motion of a spinning particle governed 
by the action (1), does not fix the value of the 
supercharge $Q_0$. We have the freedom to choose its value and 
any choice gives a consistent model [1]. On
the other hand, in Refs.[15,18], a simple exact solution of
the generalized Killing equations, corresponding to trajectories
lying on a cone, is given. Again this particular solution
requires that $\Gamma_* = 0$ and the constant of motion $J_0$ 
reduces to the standard "relative electric charge" $q$.

Finally, let us consider a solution of the homogeneous eq. (11)
for $n=1$ of the type (13). Using the Killing vectors (20) and
the Killing-Yano tensors (25) we can form the combinations:
\be
\tilde {\cal J}^{(1)}_{Aj\mu} =
R_{A\mu} f_{j \lambda\sigma}\psi^\lambda \psi^\sigma, ~~~A=0,
\dots,3; j=1,2,3.
\ee

After some algebra we get the new constants of motion of the
form (15):
\ba 
\nonumber
{\cal J}_{Aj} &=& f_{j_ \lambda\sigma}\psi^\lambda \psi^\sigma
\left( R_{A \mu}\Pi^\mu + \frac i2  R_{A [\alpha;\beta]}
\psi^\alpha \psi^\beta\right) \\
&=& -4i S_{j} J_{A} ~~~,~~~ A=0,\dots,3;j=1,2,3
\ea
where we used eqs.(15), (29-31) and (39). Strictly speaking, the
constants ${\cal J}_{Aj}$ are not completly new, being expressed
in terms of the constants $J_{A}$ and $S_{j}$. However, the
combinations (46) arise in a natural way as solutions of the
generalized Killing equations and appear only in the spinning
case. Moreover, we can form a sort of Runge-Lenz vector
involving only Grassmann components:
\be
L_{i} = \frac 1m \epsilon_{ijk} S_{j} J_{k}
~~~;~~~ i,j,k=1,2,3
\ee
with the commutation relations like in eqs.(37), (38):
\ba
\{ L_i , J_j \} &=& \epsilon_{ijk} L_{k}\\
\{ L_i , L_j \} &=& \left( \vec{S}\vec{J} - \vec{S}^2
\right)\frac{1}{m^2} \epsilon_{ijk} J_{k}.
\ea

Note also the following Dirac brackets of $L_i$ with supercharges:
\ba
\{ L_i , Q_0 \} &=& -\frac{1}{2m} \epsilon_{ijk}
Q_{j} J_{k}\\
\{ L_{i} , Q_{j} \} &=& \frac{1}{2m} \left(\epsilon_{ijk} Q_0 
J_k - \delta_{ij} Q_{k} M_{k}^{-} + Q_{i}
M_{j}^{-}\right).
\ea
\section{Concluding remarks}~~
The constants of motion of a scalar particle in a curved
spacetime are determined by the symmetries of the manifold, and
are expressible in terms of the Killing vectors and tensors,
i.e. if a spacetime admits a Killing tensor $K_{\mu_1 \dots
\mu_r}$ of valence $r$, then the quantity $K_{\mu_1\dots
\mu_r}\Pi^{\mu_1} \dots\Pi^{\mu_r}$ is conserved along the
geodesic. On the other hand, the Killing-Yano tensors can be
understood as objects generating {\it non-generic}
supersymmetries [3]. They have been also used to investigate the 
motion of spinning particles including electromagnetic interactions 
[5] and torsion [19]. The aim of this paper was to point out the 
role of the Killng-Yano tensors to generate solutions of the
homogeneous parts of the generalized Killing equations. This
solutions must be included in the complete solution of the
system of coupled Killing equations. The general procedure was
applied to the particularly case of the Taub-NUT spinning space.
The extension of this results for the motion of spinning
particles in spaces with torsion and/or in the presence of an
electromagnetic field will be discussed elsewhere.
\subsection*{Acknowledgements}~~
One of the authors ( M.V.) would like to thank A.J.Macfarlane for a
useful discussion on the properties of Killing-Yano tensors.
He has profitied from a stimulated meeting in Dubna, Russia for
which he thanks the organizers.

\end{document}